\documentclass{kapproc} 
\usepackage[dvips]{graphicx}
\upperandlowercase
\kluwerbib 

\begin{document}

\articletitle[Accretion in Cataclysmic Variable Stars]
{Accretion in Cataclysmic\\ 
Variable Stars}

\author{Brian Warner and Patrick A.~Woudt}

\altaffiltext{}{Department of Astronomy,
University of Cape Town, Rondebosch 7700, South Africa}
\email{warner@physci.uct.ac.za, pwoudt@circinus.ast.uct.ac.za}

\begin{abstract}
We consider accretion onto the white dwarfs in 
cataclysmic variables in relation to nova eruptions, dwarf nova 
outbursts, hibernation and non-radial oscillations.
\end{abstract}


The evolution of the surface temperature of an accreting white 
dwarf (WD) in a cataclysmic variable (CV) resembles a roller 
coaster. While still detached the WD cools like a single star, but as 
soon as mass transfer starts there are episodes of heating and 
cooling as either high mass transfer ($\dot{M}$) occurs through a stable 
accretion disc (i.e., as a nova-like variable) or episodically in a 
thermally unstable accretion disc (i.e., as a dwarf nova); and 
throughout its life the CV intermittently undergoes heating during 
the unstable thermonuclear burning that fuels nova eruptions.

    Although the duty cycles of dwarf novae are directly observable, 
those of novae are in general not (the exception being recurrent 
novae, which is some extremum of a continuous distribution of 
nova recurrence times). There has been a lack of agreement 
between the theoretical mass $\Delta M$ ejected in a nova eruption and 
the observed mass of nova shells. Until recently, the most elaborate 
hydrodynamic and thermonuclear models gave $\Delta M \sim  2 \times 10^{-5}$ 
M$_{\odot}$, whereas observational estimates are $\sim 2 \times 10^{-4}$
M$_{\odot}$. This discrepancy has now probably been removed by the more 
careful consideration of the thermodynamic structure of accreting white 
dwarfs (Townsley \& Bildsten 2003). Recalculation of CVs through 
entire nova cycles (as in Shara, Prialnik \& Kovetz 1993) may now 
give nova recurrence times (= $\Delta M/<\dot{M}>$) as functions of 
$<\dot{M}>$ and $M_{wd}$ which are closer to reality.

   However, the estimate of the mean value $<\dot{M}>$ between nova 
eruptions remains difficult. The great majority of novae are seen to 
erupt from a high $\dot{M}$ state and return to the same state for at 
least 100 -- 200 years after eruption (e.g. Warner 2002). Even 
though both mass and angular momentum are lost during an 
eruption, simple conservation laws require that the enhanced $\dot{M}$ 
that occurs through post-eruptive irradiation of the companion by 
the heated WD be balanced by a lengthy stage of low $\dot{M}$ 
(Kovetz, Prialnik \& Shara 1988). Whether this is enough to drive 
the CV into a state of hibernation, with low or zero $\dot{M}$ for great 
lengths of time, is disputed (e.g. Naylor 2002); it certainly does not 
seem to happen within the first couple of centuries after eruption, 
but the complete lack of success in finding the remnants of bright 
novae recorded in Oriental records of two or more millennia ago 
(Shara 1989), despite the fact that most modern naked eye novae 
end up as $m_v \sim$ 12 -- 15 remnants that are easily recognisable, 
provides the strongest evidence that hibernation on time scales 
$\sim 10^3$ y does indeed happen. Until indirect ways of estimating the 
duty cycle of this process are found, the value of $<\dot{M}>$ will 
remain very uncertain.

   Hibernating novae will appear as detached WD/M dwarf binaries 
in which the M dwarf almost fills its Roche lobe. Such detached 
systems are also responsible for the orbital period gap, according 
to the disrupted magnetic braking model (e.g. Verbunt 1984), and 
may not be distinguishable from those only temporarily at low 
$\dot{M}$ because of a nova eruption. Outside the period gap, however, 
there should be no ambiguity, and it is therefore of interest that 
there are two CVs that have recently been recognised as 
hibernating systems -- namely BPM 71214, with $P_{orb}$ = 4.84 h, and 
EC\,13471-1258 with $P_{orb}$ = 3.62 d (Kawka \& Vennes 2003; 
O'Donoghue et al.~2003). These are both relatively bright objects 
($m_v$ = 13.6 and 14.8, respectively) and would be among the 
brightest nova-like variables in the sky if they currently had $\dot{M} 
\sim 10^{-8}$ M$_{\odot}$ y$^{-1}$. 
Another object, LTT 329 at $m_v$ = 14.5, with weak 
emission indicative of extremely low $\dot{M}$, has been known for 
some time (Bragaglia et al.~1990). None of these systems are close 
to the positions of ancient observed novae (Stephenson 1986).

    Typical values of M$_{\rm v}$ before and after eruption imply 
$\dot{M} \sim 10^{-8}$ M$_{\odot}$ y$^{-1}$, 
which is an order of magnitude larger than what can be 
accounted for by magnetic braking alone; these are maintained for 
at least 200 y after eruption, but we do not know for how long 
these high rates may operate before eruption. The similar M$_{\rm v}$, and 
hence $\dot{M}$, pre- and asymptotically post-eruption has been 
explained as an irradiative feedback effect that generates an 
equilibrium high $\dot{M}$ that prevents the WD from cooling below 
$\sim$ 50\,000 K (Warner 2002). If there are $\sim$ 500 y of 
$\dot{M}$ at $\sim$ 10 times that dictated by magnetic braking (or GR 
angular momentum loss), then $\sim$ 5000 y of zero or very low $\dot{M}$ 
are required to redress the period of high living. These indicate the 
expected ratio of space densities of such systems, but such large numbers of 
low $\dot{M}$ systems were not found in an initial search for hibernating 
CVs (Shara 1989).

   However, the systematic discovery of fainter CVs (which is 
where the very low $\dot{M}$ systems inevitably will be found) has 
begun in the output of the Sloan Digital Sky Survey, the first two 
releases of which are now available (Szkody et al.~2002a, 2003). 
Based on the spectra obtained for these $\sim$ 50 new CVs, about 15\% 
have such low $\dot{M}$ values that the WD is easily detected in the 
visible spectrum. This is the type of survey that is needed to find 
very low $\dot{M}$ systems -- at least for CVs with short orbital periods 
where both the WD and its companion are intrinsically faint (very 
low $\dot{M}$ systems of longer $P_{orb}$ may be harder to recognise in the 
initial colour survey because they will look like ordinary K or M 
red dwarfs). With more complete surveys to even fainter limits it 
should become possible to estimate the true frequency distribution 
of $\dot{M}$, from which the $\dot{M}$ duty cycle between nova eruptions 
will follow.

      Sion (2003) has shown that the observed surface temperatures 
$T_{eff}$ of the WDs in dwarf novae can be fully accounted for by the 
compressional heating that accompanies accretion. For the short 
orbital periods (i.e., below the orbital period gap) $T_{eff}$ clusters 
around 15\,000 K, which is close to what is predicted for GR-driven 
evolution. This result depends somewhat on the adopted WD 
masses. One importance of this result is that $T_{eff}$ is an indirect way 
of learning something about $<\dot{M}>$ and the $\dot{M}$ duty cycle for 
dwarf novae.

    The observed clustering of $T_{eff}$ around 15\,000 K, seen in the list 
given in Table 1 (based on Sion 2003), contains an observational 
bias -- the stars on which it is based have almost all been found 
from dwarf nova outbursts, which have intervals that become very 
long for very low $\dot{M}$. For example, in Table 1 the lowest 
observed temperatures are correlated with the greatest intervals, 
$T_{out}$, between outbursts\footnote{WZ Sge appears not to fit this 
correlation, but it has a magnetic primary, which is probably the reason for 
the large $T_{out}$ (e.g. Warner, Livio \& Tout  1992).}. 
There should be other, lower $T_{eff}$ systems, 
with outburst intervals so long that they are unlikely to have been 
found via outbursts and are therefore absent from studies made 
hitherto. But these are the intrinsically faint CVs that are beginning 
to be found spectroscopically in the Sloan Survey.

   This point appears most strongly in the known CV WD primaries 
that have non-radial oscillations. The ZZ Cet instability strip for 
isolated WDs lies approximately in the range 11\,000 -- 12\,000 K. It 
may be modified to some extent in accreting WDs where the outer 
envelope has a different physical and chemical structure (see, e.g., 
Townsley \& Bildsten 2003). Until recently the only known CV/ZZ 
combination was GW Lib\footnote{This ZZ Cet star is overlooked in 
the total given by Fontaine et al.~(2002) and Bergeron et al.~(2003).}
(Warner \& van Zyl 1998), which has $T_{eff}$ = 14\,700 K according to 
Szkody et al.~(2002b), indicating that 
the instability strip may be displaced blueward of that for isolated 
WDs. Note, however, that there are several dwarf novae in Table 1 
that have measured $T_{eff}$ similar to that of GW Lib, have detectable 
WD absorption lines in their spectra, and have been sufficiently 
observed photometrically in quiescence for ZZ Cet type 
oscillations to have been detected -- without success. But they 
typically have outburst intervals $\sim$ 1 y, which may be too short a 
time in quiescence for oscillations to grow in\footnote{Growth times 
for non-radial oscillations in ZZ Cet stars can be anywhere from 
hours to thousands of years, according to which mode is being 
excited (Goldreich \& Wu 1999).}. GW Lib, on the 
other hand, has had only one known outburst (in 1983: by which it 
was identified as a CV); this incidentally demonstrates that 
oscillations can appear within less than a decade after outburst.

   The next CV/ZZ to be discovered was SDSS1610 (Woudt \& Warner 2003), 
which was identified in the Sloan Survey first 
release as a very low $\dot{M}$ system, and has had no known 
outbursts. On the basis of these two examples alone we are led to 
suspect that CV/ZZ stars are extremely low $\dot{M}$ systems and their 
$T_{eff}$ will be found to be lower than that currently accepted for GW 
Lib (and which in any case may have been overestimated). 

    Once well calibrated, the relative frequency of CV/ZZ systems 
will provide a valuable indicator of the number of CVs in the 
accreting WD instability strip, which is another means of studying 
the $\dot{M}$ history of CVs. The first Sloan Survey release had just 
one CV/ZZ (SDSS\,1610) out of 25 objects. The second release 
(Szkody et al.~2003) has four or five potential candidates out of 35 
candidates, based on the visibility of the WD absorption spectra, 
but with only two of these having spectra very closely similar to 
GW Lib or SDSS1610 (i.e., rather than like the non-oscillating 
systems such as Z Cha and OY Car, which have more emission-
filled absorption lines). Our observations (Warner \& Woudt  
2003) nevertheless show that SDSS\,0131 and SDSS\,2205 are 
certainly CV/ZZ stars -- the fifth candidate has yet to be observed. 
A frequency $\sim$ 8\% among faint CVs is therefore indicated, which 
will give $\sim$ 32 systems once the estimated 400 new CVs expected 
in the Sloan Survey (Szkody et al.~2002a) have been found and 
interrogated photometrically.

    The four known CV/ZZ stars have hydrogen emission cores 
superimposed on the WD absorption lines. For somewhat lower 
$\dot{M}$ the emission lines would not be so readily observable. It 
would be worth searching carefully for weak emission cores in 
known ZZ Cet stars -- there could be a hibernating CV hidden 
among them.

\begin{table}[ht]
\caption[White Dwarf temperatures in selected CVs.]
{White Dwarf temperatures in selected CVs.}
\begin{tabular*}{\textwidth}{@{\extracolsep{\fill}}lrrrl}
\sphline
\it Star &\it $P_{orb}$ (min) &\it $T_{eff}$ (K)& \it $T_{out}$ (d)& \it References \cr
\sphline
{\bf Non-magnetic} & & & & \cr
GW Lib   &  76.8   &     13\,300  & $>$7000   &  Szkody et al.~2002b \cr
BW Scl  &   78.2   &      14\,800 &           &    Szkody et al.~2002c  \cr
LL And  &   79.8   &      14\,300 &    5000:  &   Howell et al.~2002a  \cr
AL Com  &   81.6   &      16\,300 &     325   &   Szkody et al.~2003  \cr
WZ Sge  &   81.6   &      15\,000 &   10000:  &   Sion et al.~1995a  \cr
SW UMa  &   81.8   &      14\,000 &     954   &   Gaensicke \& Koester 1999 \cr
HV Vir  &   83.5   &      13\,300 &    3500:  &   Szkody et al.~2002c  \cr
WX Cet  &   84.0   &      13\,000 &    1000:  &   Sion et al.~2003  \cr
EG Cnc  &   86.3   &      12\,300 &    7000:  &   Szkody et al.~2002c  \cr
BC UMa  &   90.1   &      15\,200 &    1500:  &   Szkody et al.~2002c  \cr
EK TrA  &   90.5   &      18\,800 &     230   &   Gaensicke et al.~2001  \cr
VY Aqr  &   90.8   &      13\,500 &     500:  &   Sion et al.~2003  \cr
OY Car  &   90.9   &      16\,000 &     160   &   Horne et al.~1994  \cr
HT Cas  &  106.1   &      15\,500 &     166:  &   Wood et al.~1992  \cr
VW Hyi  &  107.0   &      22\,000 &      28   &   Sion et al.~1995b  \cr
Z Cha   &  107.3   &      15\,700 &      51   &   Robinson et al.~1995  \cr
CU Vel  &  113.0   &      15\,000 &     165   &   Gaensicke \& Koester 1999 \cr
EF Peg  &   123    &      16\,600 &     250:  &   Howell et al.~2002a  \cr
 & & & & \cr
{\bf Polars} & & & & \cr
EF Eri   &   81.0   &       9\,500  &           &     Howell et al.~2002b\cr
DP Leo   &   89.8   &      13\,500  &           &     Schwope et al.~2002\cr
VV Pup   &  100.4   &       9\,000  &           &     Skzody et al.~1983\cr
V834 Cen &  101.5   &      15\,000  &           &     Beuremann et al.~1990\cr
BL Hyi   &  113.7   &      20\,000  &           &     Wickramasinghe et al.~1984\cr
ST LMi   &  113.9   &      11\,000  &           &     Mukai \& Charles 1986\cr
MR Ser   &  113.6   &      $<$8\,500  &           &     Schwope \& Beuremann 1993\cr
AN UMa   &  114.8   &     $<$20\,000  &           &     Sion 1991\cr
HU Aqr   &  125.0   &     $<$13\,000  &           &     Glenn et al.~1994\cr
UZ For   &  126.5   &      11\,000  &           &     Bailey \& Cropper 1991\cr
AM Her   &  185.7   & $\sim$20\,000  &           &     De Pasquale \& Sion 2001\cr
RXJ1313  &  251.4   &      15\,000  &           &     Gaensicke et al.~2000\cr
\sphline
\end{tabular*}
\begin{tablenotes}
$:$ Uncertain value\newline
\end{tablenotes}
\end{table}
\inxx{captions,table}

    Table 1 also includes $T_{eff}$ measurements for polars, i.e., for the 
strongly magnetic WDs in CVs. Several of these have lower 
temperatures than any seen in dwarf novae. Some have $T_{eff}$ that 
could put them in the instability strip. Most of these are the 
brightest and best-observed polars, but none have been found to 
have ZZ Cet behaviour. Theoretical investigations are needed that 
include the effects of strong fields, which will presumably be 
found to prevent non-radial oscillations above some critical field 
strength.

    The low $T_{eff}$ in the longer $P_{orb}$ polar RXJ1313 implies a much 
lower $<\dot{M}>$ than is the case for the commonly observed non-magnetic 
CVs at that $P_{orb}$. Again there may be an observational 
bias in action -- polars even of low $\dot{M}$ are easily found through 
their hard X-Ray emission, whereas the comparative rarity of low 
$\dot{M}$ dwarf novae near $P_{orb}$ of 4 h is probably caused by the effect 
of irradiation-enhanced $\dot{M}$, which turns them either into high 
$\dot{M}$ nova-likes, or into extremely low $\dot{M}$ dwarf novae (or even 
deeply hibernating, essentially zero $\dot{M}$ systems, as in the BPM 
and EC objects discussed above) that are hard to find (Wu, 
Wickramasinghe \& Warner 1995).

    Finally, we draw attention to the AM CVn (helium-transferring 
CV) systems, where the WD primaries could in principle show 
non-radial oscillations if they are in the equivalent of the DB 
variable instability strip. These would have to be found among AM 
CVn stars that have an appropriate $\dot{M}$ -- but that is in just the 
range where these systems show VY Scl behaviour, and in the high 
state the accretion luminosity will overwhelm any intrinsic 
pulsations of the primary, while in the low state $\dot{M}$ is probably 
too variable to give the oscillations time to grow.

\begin{acknowledgments}
    BW's research is funded by the University of Cape Town; 
PAW's research is funded by a strategic grant from the University 
to BW and by funds from the National Research Foundation.
\end{acknowledgments}


\begin{chapthebibliography}{1}
\bibitem{ba91}  Bailey, J. \& Cropper, M. (1991). MNRAS, 253, 27
\bibitem{be03}  Bergeron, P., Fontaine, G., Billeres, M., Boudreault, S. \& Green, 
                E.M. (2003). ApJ, in press
\bibitem{be90}  Beuermann, K., Schwope, A.D., Thomas, H.-C. \& Jordan, S. 
                (1990). In Accretion Powered Compact Binaries, ed. C.W. Mauche, 
                Cambridge University Press, 265
\bibitem{br90}  Bragaglia, A., Greggio, L., Renzini, A. \& D'Odorico, S. (1990). 
                ApJL, 365, 13
\bibitem{de01}  De Pasquale, J. \& Sion, E.M. (2001). ApJ, 557, 978
\bibitem{fo03}  Fontaine, G., Bergeron, P, Bill\`eres, M. \& Charpinet, S. (2003). ApJ, 591, 1184
\bibitem{gae99} Gaensicke, B.T, \& Koester, D. (1999). A\&{A}, 346, 151
\bibitem{gae00} Gaensicke, B.T., Beuermann, K., De Martino, D. \& Thomas, H.-C. (2000). A\&{A}, 354, 605
\bibitem{gae01} Gaensicke, B.T., Szkody, P., Sion, E.M., Hoard, D.W., Howell, S., Cheng, F.H. \& 
                Hubeny, I. (2001). A\&{A}, 374, 656
\bibitem{gl94}  Glenn, J., et al. (1994). ApJ, 424, 967
\bibitem{go99}  Goldreich, P \& Wu, Y. (1999). ApJ, 511, 904
\bibitem{ho94}  Horne, K., Marsh, T.R., Cheng, F.R., Hubeny, I. \& Lanz, T. (1994). ApJ, 426, 294
\bibitem{ho02a} Howell, S.B., Gaensicke, B.T., Szkody, P. \& Sion, E.M. (2002a). ApJL, 575, 419
\bibitem{ho02b} Howell, S.B., Harrison, T.E. \& Osborne, H. (2002b). AAS Abstract, 201, 4008
\bibitem{ka03}  Kawka, A. \& Vennes, S. (2003). These Proceedings
\bibitem{ko88}  Kovetz, A., Prialnik, D. \& Shara, M.M. (1988). ApJ, 325, 828
\bibitem{mu86}  Mukai, K. \& Charles, P.A. (1986). MNRAS, 222, 1P
\bibitem{na02}  Naylor, T. (2002). In Classical Nova Explosions, eds M. Hernanz \& 
                J. Jose, AIP Conf. Proc., 637, 16
\bibitem{dod03} O'Donoghue, D., Koen, C., Kilkenny, D., Stobie, R.S., Koester, D., Bessell, M.S.,
                Hambly, N. \& MacGillivray, H. (2003). MNRAS, in press
\bibitem{ro95}  Robinson, E.L., et al. (1995). ApJ, 443, 295
\bibitem{sc93}  Schwope, A.D. \& Beuermann, K. (1993). In White Dwarfs: Advances in Observation \& Theory, 
                ed. M.A. Barstow, Kluwer, 381
\bibitem{sc02}  Schwope, A.D., Hambaryan, V., Schwarz, R., Kanbach, G. \& Gaensicke, B.T. (2002). 
                A\&{A}, 392, 541
\bibitem{sh89}  Shara, M.M. (1989). PASP, 101, 5
\bibitem{sh93}  Shara, M.M., Prialnik, D. \& Kovetz, A. (1993). ApJ, 406, 220
\bibitem{si91}  Sion, E.M. (1991). AJ, 102, 295.
\bibitem{si03}  Sion, E.M. (2003). In Proc. 13th European Workshop on White Dwarfs (Naples), in press
\bibitem{si95a} Sion, E.M., et al. (1995a). ApJ, 439, 957
\bibitem{si95b} Sion, E.M., Szkody, P., Cheng, F. \& Min, H. (1995b). ApJL, 444, 97
\bibitem{sio03} Sion, E.M., Szkody, P., Cheng, F., Gaensicke, B.T. \& Howell, S.B. (2003). ApJ, 583, 907
\bibitem{st86}  Stephenson, R.F. (1986). In RS Ophuichi and the Recurrent Nova Phenomenon, ed. M.F. Bode, 
                VNU Sci. Press (Utrecht), 105
\bibitem{szk83} Szkody, P., Bailey, J.A. \& Hough, J.H. (1983). MNRAS, 203, 749
\bibitem{sz02a} Szkody, P., et al. (2002a). AJ, 123, 430
\bibitem{sz02b} Szkody, P., Gaensicke, B.T., Howell, S.B. \& Sion, E.M. (2002b). ApJ, 575, 79
\bibitem{sz02c} Szkody, P., Sion. E.M., Gaensicke, B.T. \& Howell, S.B. (2002c). ASP Conf. Ser., 261, 21
\bibitem{sz03}  Szkody, P., et al. (2003). AJ, 126, 1499
\bibitem{tb03}  Townsley, D.M. \& Bildsten, L. (2003). ApJ, in press
\bibitem{ver84} Verbunt, F. (1984). MNRAS, 209, 227
\bibitem{wa02}  Warner B., (2002). In Classical Nova Explosions, eds M. Hernanz \& 
                J. Jose, AIP Conf. Proc., 637, 3
\bibitem{wa98}  Warner, B. \& van Zyl, L. (1998). IAU Symp. No. 185, 321
\bibitem{wo03b} Warner, B. \& Woudt, P.A. (2003). Proc. IAU Colloq. 193, in press
\bibitem{wlt92} Warner, B., Livio, M. \& Tout, C.A. (1992). MNRAS, 282, 735
\bibitem{wi84}  Wickramasinghe, D.T., Visvanathan, N. \& Tuohy, I.R. (1984). ApJ, 286, 328
\bibitem{whv92} Wood, J.H., Horne, K. \& Vennes, S. (1992). ApJ, 385, 294
\bibitem{wo03a} Woudt, P.A. \& Warner, B. (2003). MNRAS, in press
\end{chapthebibliography}
\end{document}